# 'DANCING WITH THE STARS': ASTRONOMY AND MUSIC IN THE TORRES STRAIT

Duane W. Hamacher, Alo Tapim,
Segar Passi and John Barsa

**ABSTRACT**: Song and dance are a traditional means of strengthening culture and passing knowledge to successive generations in the Torres Strait of northeastern Australia. Dances incorporate a range of apparatuses to enhance the performance, such as dance machines (*Zamiyakal*) and headdresses (*Dhari*). The dances, songs, headdresses and dance machines work together to transfer important knowledge about subsistence survival, social structure, and cultural continuity. This paper explores how celestial phenomena inspire and inform music and dance.

## Introduction

The Torres Strait is an archipelago of approximately 275 islands stretching between Cape York Peninsula on mainland Australia and Papua New Guinea, of which 17 are inhabited by communities. The islands are grouped into five major regions: eastern, central, northern, western, and southwestern (Fig. x.1).

Torres Strait Islanders are a Melanesian people indigenous to Australia.[1] As a seafaring people, their traditions contain a significant astronomical component. This is reflected in their traditional laws, creation stories, ceremonies, and daily activities. It is also reflected in the Islander flag (Fig. x.1 inset) as a five-pointed star. The five points represent the Strait's five major island groups, and the star itself represents an unspecified navigational star. Two major indigenous languages are spoken: the Papuan *Meriam Mir* language in the eastern islands, and the Austronesian *Kala Lagaw Ya language* (and various dialects) in the central, northern, western, and southwestern islands. Each of the five major island groups have distinct customs, but share commonalities in culture and tradition.

Much of the early ethnographic literature on Islander cultures comes from the London Missionary Society, which established missions in the Torres Strait in the 1840s, and two major ethnographic expeditions from Cambridge University, led by Alfred Cort Haddon in 1888 and 1898.[2] The Haddon expeditions focused on Mer and Mabuyag islands.

---

[1] Nonie Sharp, *Stars of Tagai* (Canberra: Aboriginal Studies Press, 1993).
[2] Alfred Cort Haddon, *Cambridge anthropological expedition to the Torres Straits*, 6 vols (Cambridge: Cambridge University Press, 1901-1935), Vol. 3.

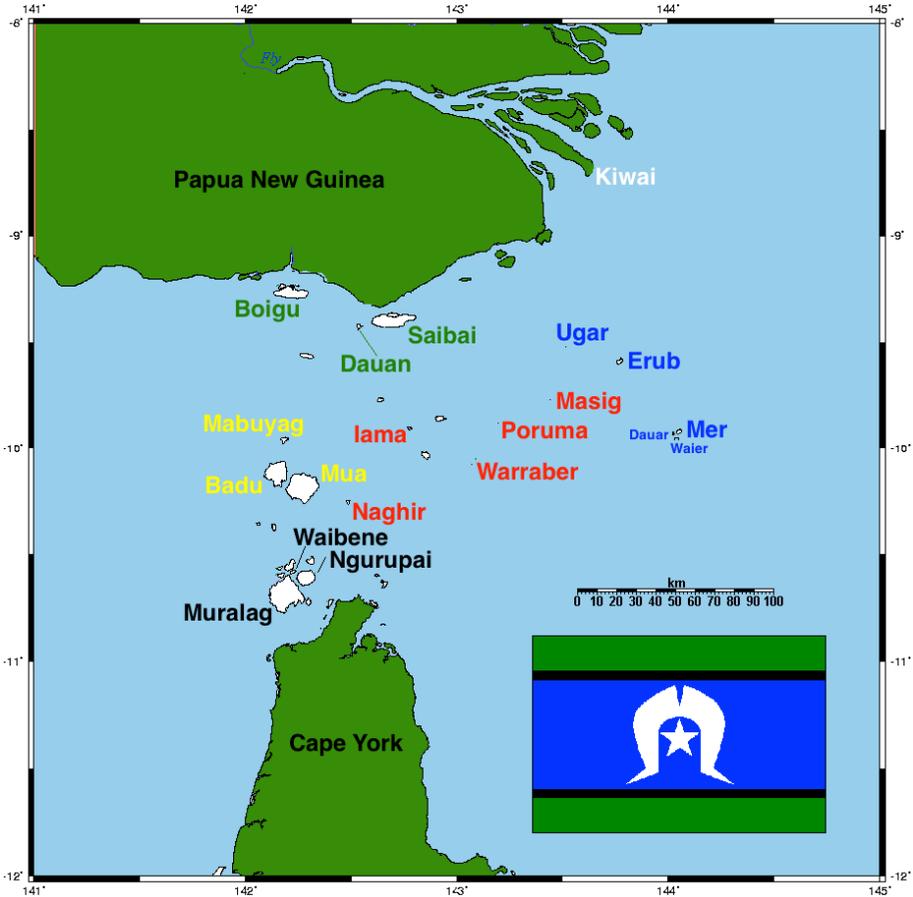

Fig. x.1: Map of the Torres Strait with labels on inhabited islands. Eastern (blue), central (red), northern (green), western (yellow), southwestern (black). Image: Wikipedia. Inset: The Torres Strait Islander flag, developed by Bernard Namok of Waiben in 1992. Green represents the landmasses of Australia and PNG, blue represents the ocean, black represents the people, the headdress is a dhari, and the star represents a navigational star, with each point representing the five major island groups. The colour white represents peace and the Islander's conversion to Christianity.

## Astronomy in Torres Strait Traditions

In Islander cultures, ways of being and ways of knowing are closely linked to the stars.[3] This relationship encompasses a sense of identity and belonging to their environment, to the peoples' understanding of themselves, and to their culture and traditions.[4] This sense of belonging links the past, present, and future into a holistic system of knowledge that developed over thousands of years. This knowledge informs Islander traditions: the laws, customs, and practices that are recorded and handed down in the form of story, song, dance, ceremony, art forms, and material culture.[5]

Islander astronomical knowledge also contains practical information about the natural world and encompasses how Islanders perceive and understand the environment in which they live.[6] This is essential for survival and cultural continuity, and is integral to the everyday lives of Islander people. Islander understandings of identity, for instance, are linked to Tagai – the creation deity that is represented by a constellation of stars that spans across the sky.[7]

The story of Tagai (aka *Thoegay*) varies across the Strait, but the general theme is that Tagai was a great warrior and fisherman. He had a crew of twelve men called *Zugubals* and his confidant and 'first mate', *Kareg*. The group went fishing on a reef but were unsuccessful in catching any fish. Tagai left to search the reef for a more suitable spot. While he was gone, the Zugubals grew hungry, tired, and frustrated in the heat. They ate all of their food and drank all of their water. They foolishly decided to consume Tagai's food and water. When Tagai returned, he was furious. He tied the men into two groups of six and cast them into the sea, where they drowned. They went into the sky as two groups of stars: the *Usiam* (the Pleiades) and *Utimal/Seg* (belt and scabbard stars of Orion).

Tagai and Kareg went to the other side of the sky to keep away from the Zugubals (this is notably similar to the Greek traditions about Orion and Scorpius being placed on opposite sides of the sky by the gods to keep them away from each other after a battle).

Tagai is a very large constellation spanning across the sky (Fig. x.2).[8] His left hand is the Southern Cross (Crux) holding a spear. His right hand is the Western constellation Corvus holding a species of Eugenia fruit. His head and body consist of stars in Centaurus and Lupus. He is standing on the bow of his canoe, traced out by the body of Scorpius. His first-mate, Kareg, is the star Antares at the stern of the canoe. Their appearance in the morning and evening throughout the year inform seasonal change, food economics, and social structure.

---

[3] Sharp, *Stars of Tagai.*
[4] Martin Nakata, personal communication, date.
[5] Jeremy Beckett, *Torres Strait Islanders: custom and colonialism* (Cambridge: Cambridge University Press, 1987).
[6] Martin Nakata, 'The cultural interface of Islander and Scientific knowledge', *Australian Journal of Indigenous Education* 39 (Supplement) (2010): pp. 53-7.
[7] Sharp, *Stars of Tagai.*
[8] Peter Eseli, *Eseli's Notebook*, ed. Anna Shnukal and Rod Mitchell (University of Queensland: Aboriginal and Torres Strait Islander Studies Unit Research Report Series, Vol. 3, 1998), p. 66.

Tagai's influence on Islander social structure encompasses four themes for governing the Islander way of life.[*] The first links the stars of Tagai as custodians of knowledge and spirituality for future generations of Islanders. The second is that Islanders are a sea-faring people who share a common way of life. The third relates to the order of the world (the laws and customs) that are instructed by Tagai. The fourth discusses the cycle of life as a period of time and renewal based on the Sun, the Moon, and the rising and setting of particular stars.

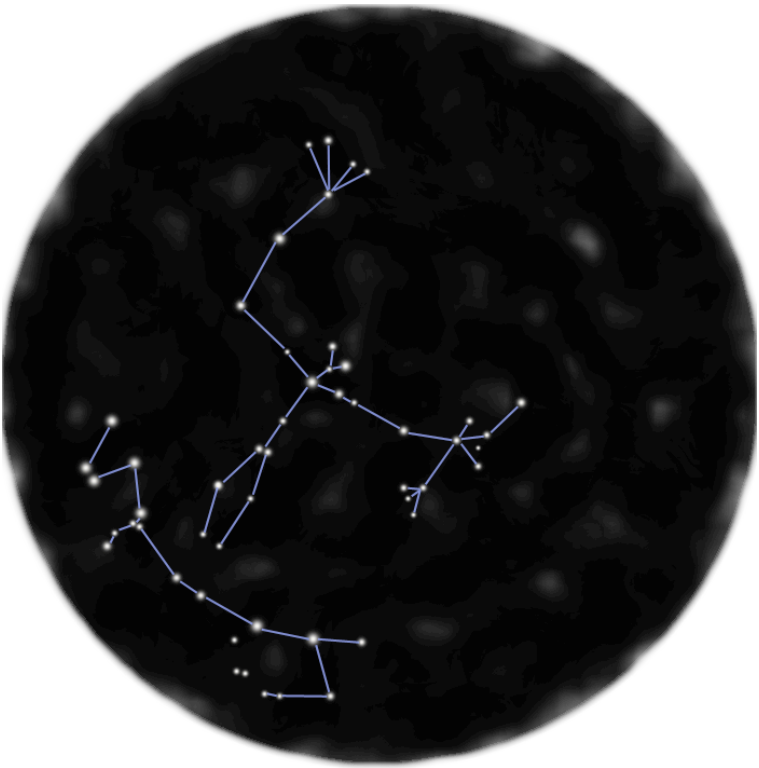

*Fig. x.2: The stars of Tagai, spanning across the night sky. This image is used as the logo for the Tagai State College primary and secondary schools in the Torres Strait. Image: Wikipedia.*

---

[*] Sharp, *Stars of Tagai*.

Thus, the night sky was recruited, structured, and woven into oral traditions (including song and dance) informing elements of the peoples' morals and values.[10] This is reflected in Malo's Laws. These are the sacred laws followed by all Meriam Islanders. One of Malo's laws is that all stars follow their own path, and so must each person. Do not follow the paths of others, but only of your own.

**Music and Dance**

Songs and dances are a way of recording knowledge and transmitting it to successive generations. The dance movements are one of the ways in which information is transmitted, and masks, headdresses, dance machines, and other props are an important element to these performances. The instruments used include the *warup* (traditional drum), bamboo drums, or contemporary instruments such as guitars. A warup is an hourglass-shaped drum carved from hollowed driftwood, using the skin of a *goanna* as the drum head.

Mua artist David Bosun created a linocut artwork in 2007 describing the important relationship between Torres Strait Islander astronomers (*Zugubau Mabaig* in western dialects) and the warup (Fig. x.3).[11] It is through the warup that these sacred teachings are echoed. The linocut features a Zugubau Mabaig holding two stars in front of a warup, echoing the sacred knowledge. Zugubau Mabaig are taught to read the stars in a *kwod* (a sacred meeting and learning house).

During a visit to Mer in July 2015, Meriam elders and dancers, including co-authors Alo Tapim and John Barsa, performed a dance called *Gedge Togia*. The music was played on two warups, with the lyrics 'Gedge Togia Malpanuka' sung by elders. Alo Tapim explained the meaning behind the song. 'Gedge togia' means 'rising over home' (in this case Mer) in the Meriam Mir language. *Malpanuka* refers to the moon in the Mabuyag dialect of the Kala Lagaw Ya language. During the Haddon expedition in the late nineteenth century, vocabulary from Mer and Mabuyag were recorded by linguist Sidney Ray.[12] He recorded the Mabuyag name for the moon as *mulpal*, which referred to the full moon (or near full moon) as opposed to the crescent moon (*inur-dan*). Mer lies 200 km across the sea, at nearly due east, from Mabuyag.

---

[10] John Whop, 'Stories Under Tagai', (video) (Brisbane: State Library of Queensland, 2012), at https://www.youtube.com/watch?v=5kU4EvV9yI8 [Accessed 4 February 2016].

[11] David Bosun, *Zugubau Mabaig*, Australian Art Network (2007a), at http://australianartnetwork.com.au/shop/artwork/zugubau-mabaig/ [Accessed 4 Feb 2016],

[12] Haddon, *Cambridge anthropological expedition*, Vol. 3, p. 152.

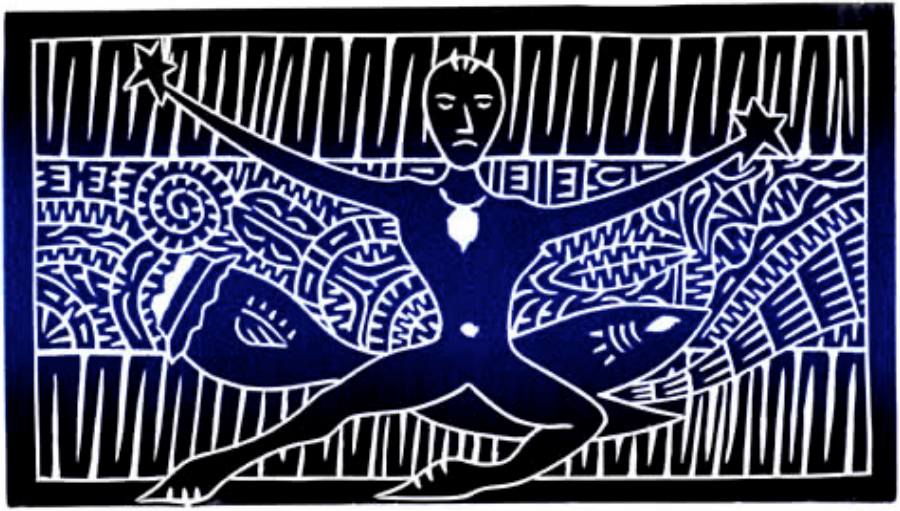

*Fig. x.3: A linocut entitled 'Zugubau Mabaig (Astronomer)' by Mua artist David Bosun.*

In the Meriam Mir language, for comparison, the moon is called *meb* (which is also the term for a month).[13] The new moon (thin crescent) is called *aketi meb*, the first quarter moon is *meb degemli*, a nearly full moon (waxing or waning gibbous) is *eip meb*, a full moon is *giz meb*, a third quarter moon is *meb zizimi,* and a lunar eclipse is *meb dimdi*. Meriam elder and co-author Segar Passi gave the name *kerkar meb*, which literally translates to 'new moon' – when it first appears as a waxing crescent.

The planting of gardens is associated with lunar phases and the moon's effect on ocean tides.[14] Vegetables are planted during the rising tide; when the tide drops, the gardener stops. Vegetables that bear edible fruit are planted during the new moon, while tubers are planted during the full moon. For example, gardeners plant yams facing the rising full moon.

Alo Tapim explained that quarter moons are more ideal for fishing than the new or full moon. This is due to the neap tides being low in amplitude, thus not churning up sand and soil that clouds the water (which was confirmed by other community members). This was confirmed by several Islander people. However, hunting crayfish and turtle are more appropriate during the full moon, when the strong moonlight (*meb gerib*) illuminates the water, making the animals easier to see, as noted by co-author John Barsa.

The *Gedge Togia* dancers hold a rod in each hand with a moon at the end: one is of the full moon and the other is of the new moon (the 'new moon' in this context colloquially describes the waxing or waning crescent when the moon is very thinly lit rather than the 'invisible' duration of the new moon familiar to Western astronomers). In addition to passing on important knowledge, Tapim explained how this song is important for demonstrating connections between different islands in the Torres Strait. During a legal battle for sea rights the Australian government in 2005, the opposition lawyer claimed that the islands were all separate enclaves with little or no contact with each other. Tapim sang this traditional song, comprising lyrics in two languages, to the judge as evidence this was incorrect. The Islanders won their case.

---

[13] Haddon, *Cambridge anthropological expedition*, Vol. 3, p. 152.
[14] Sharp, *Stars of Tagai*, p. 60.

## Headdresses

Headdresses (*dhari* in Meriam Mir and *dhoeri* in Kala Lagaw Ya) are prominent in Islander dance and ceremony, particularly the *kab kar* (sacred traditional dance). Traditional masks are made from a variety of components, including feathers, wood, and shells. Some headdresses are large and elaborate, representing various things such as animals, stars, and vehicles. Although worn by warriors in pre-colonial times, they are almost exclusively worn during dances and ceremonies today. Dhari are a common representation of Torres Strait culture, which is why they are featured on the Islander flag.

A *Madthubau Dhibal* headdress (Fig. x.4, right), created by Jeff and Sedrick Waia from Saibai, is associated with a rain dance and the constellations that comprise Thoegay (Tagai) and *baidam* (the shark). A man is chosen from each clan to watch the night skies for the movements of these constellations. The constellations inform seasonal change, animal behaviour, and gardening. The dance welcomes the coming monsoon season (*Kuki*), which lasts from January to April.

Knowledge of Thoegay and other constellations by the Islander astronomer is essential for food production, especially when preparing for the Kuki. Thoegay/Tagai and the Zugubals he cast into the sky are important for denoting seasonal change. Mabuyag elder Peter Eseli explains that because Thoegay placed the Zugubals on the opposite side of the sky from he and Kang (Kareg), the Zugubals must announce their appearance in the dusk sky with thunder.[15] The dusk rising of Usiam (Pleiades) occurs in mid-November.[16] By early December, both Usiam and Utimal/Seg (Orion) rise at dusk. This coincides with the coming monsoon rains, accompanied by the sound of thunder. At this time, Thoegay and Kang dive into the sea as they set below the horizon, splashing water into the sky. This water causes the first rains of the Kuki.[17]

---

[15] Eseli, *Eseli's Notebook*, p. 17.
[16] When the stars have an altitude of 5° and the Sun −16°. See Trevor M. Leaman, Duane W. Hamacher, and Mark Carter. 'Aboriginal Astronomical Traditions from Ooldea, South Australia, Part 2: Animals in the Ooldean Sky', *Journal of Astronomical History and Heritage* 19, no. 1 (2016): pp. 61–78.

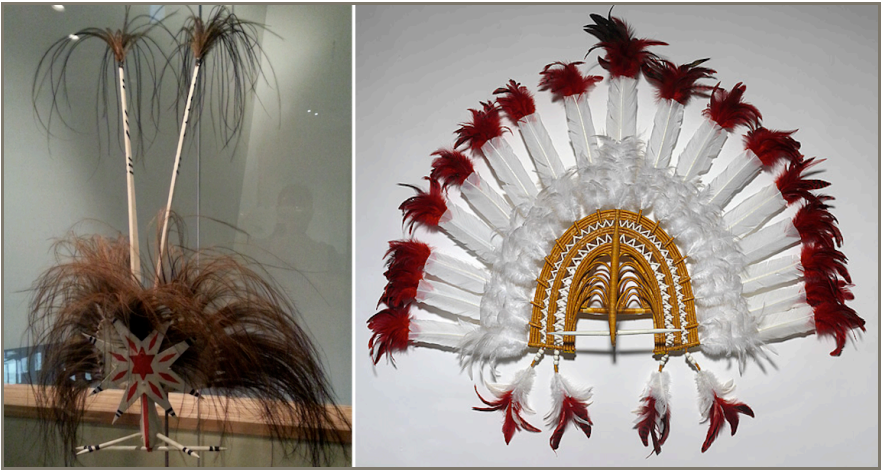

Fig. x.4: (Left) An eclipse mask by Sipau Gibuma from Boigu ca. 1990. Photo by D. W. Hamacher at the National Museum of Australia, Canberra. (Right) Madthubau Dhibal headdress by Jeff Waia from Saibai ca. 2008. National Gallery of Australia, Accession No: NGA 2009.821.

The *Mayngu dhoeri* (Fig. x.4, left) was created by Sipau Audi Gibuma from Boigu. It is used on very special occasions for a performance called *meripal kulkan patan*, meaning 'blood covering the moon'. This performance is conducted during a total lunar eclipse. In Islander traditions, a total lunar eclipse was an omen of war and bloodshed from another island or the Papuan mainland.[18] During the dance, the names of all the islands in the Torres Strait are chanted in a cycle. The island named when the moon emerges from totality is the island from which the threat of war is imminent.

Mua artist David Bosun refers to a lunar eclipse as *merlpal maru pathanu*. According to Bosun, this means 'the ghost has taken the spirit of the moon' (from mari/maru = ghost/spirit, and mulpal/merlpal = moon).[19] During a lunar eclipse, warriors prepare for battle while women and children hide.

---

[17] Sharp, *Stars of Tagai*, pp. 60-1.
[18] Lindsay Wilson, *Kerkar Lu: contemporary artefacts of the Torres Strait Islanders* (Brisbane: Queensland Department of Education, 1993), p. 104.
[19] David Bosun, *Merlpal Mari Pathanu*. Australian Art Network (2007b), available at http://australianartnetwork.com.au/shop/artwork/merlpal-mari-pathanu/ [accessed 4 February 2015].

Bosun says the term can be used for both lunar and solar eclipses. A lunar eclipse in the Meriam Mir language is called *meb dimdi*, meaning 'covered moon'.[20]

Lunar eclipses occur about once every 2.5 years from any given location on Earth. The duration of totality can last up to 107 minutes.[21] The relationship between blood and the reddening of the moon during a total lunar eclipse is found in many indigenous traditions. On mainland Australia, a lunar eclipse was seen by many Aboriginal groups as the moon-man covered in blood and was linked to death and omens.[22]

**Dance Machines**

Dance machines (*zamiyakal*) are dynamic mechanical devices used by dancers to enhance performance and are prominent in Islander dances of the western, northern, and central islands (they are not common in the eastern islands). Zamiyakal serve as a mnemonic for recording and sustaining knowledge and cultural traditions.[23] The machines can mimic the motions of celestial bodies, such as the rising of *Ilwel* (Venus), the movement of a bright meteor across the sky, or the setting of the Southern Cross.

Star designs utilising various mechanical motions are incorporated into dance machines. Some are folded in half, which the dancer expands to reveal, then closes to hide. Some are contracted at the end of bamboo poles. An example is the *Titui* (star) machine from Mua (Fig. x.5A). The underside is pushed forward by the dancer, extending the rays of the star outward like an inverted umbrella. Other star designs are an ornamental component of a zamiyakal.

The *Koey Thitui Migi Thithui* dance from St Paul's community on Mua signifies the importance of the Southern Cross (left hand of Thoegay) and its use in navigation.[24] John Barsa notes that Crux is used as a compass, as it points southward. Similarly, a *Southern Cross* machine, developed by Ian Larry of Warraber[25] (Fig. x.5B), demonstrates the constellation's importance for denoting seasonal change. The song describes the relationship between the setting constellation and the north winds of the *Nay Gay* season from October to December. The Southern Cross sets at dusk in October, during which time gardeners begin preparing for the coming Kuki. Grass is burned then dug into the soil for planting yams, sugar cane, and banana.[26] This also relates to *The Tagai* machine, made by Tabipa Harry of Dauan (Fig. x.5C). It mimics the movement of Tagai is he casts his spear into the sea (held in his left hand, the Southern Cross).

---

[20] Haddon, *Cambridge anthropological expedition,* Vol. 3, p. 152.
[21] Karttunen, Hannu, Pekka Kröger, Heikki Oja, Markku Poutanen, and Karl J. Donner. *Fundamental Astronomy.* (Berlin, Heidelberg, New York: Springer, 2007), p. 139.
[22] Duane W. Hamacher and Ray P. Norris, 'Eclipses in Australian Aboriginal Astronomy', *Journal of Astronomical History and Heritage* 14, no. 2 (2011): pp. 103-114.
[23] John T. Kris in Robyn Fernandez and Emma Loban, *Zamiyakal: Torres Strait dance machines* (Thursday Island: Gab Titui Cultural Centre, 2009), p. 4.
[24] Will Kepa, Nigel Pegrum, and Karl Nuenfeldt. *Lagau Kompass: Music & Dance From St. Paul's Community (Mua Island) Torres Strait* (CD/DVD) (Cairns: Pegasus Studios, 2013).
[25] Fernandez and Loban, *Zamiyakal,* p. 42

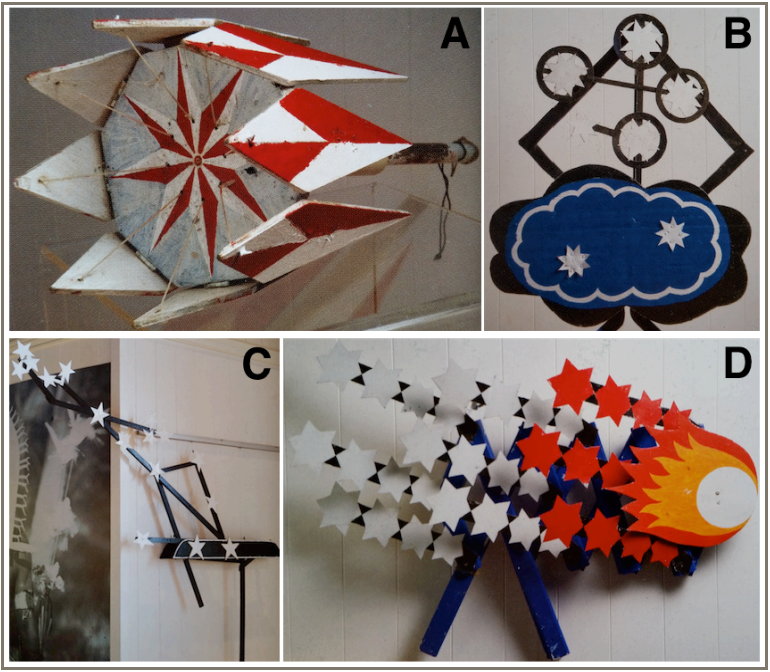

Fig. x.5: Astronomical dance machines (A) Titui, (B) Southern Cross, (C) The Tagai, (D) Comet.[27]

Cultural traditions are also encoded in dance. The *Comet* machine from Iama (Fig. x.5D) utilises a mechanical movement that makes it appear to 'shoot' across the sky. This describes a bright meteor rather than a comet (the two are often conflated in language translations). Alo Tapim explains that in Meriam traditions, bright meteors (fireballs) are called *Maier*. They signify that an important person has passed away (a common association across the

---

[26] Eseli, *Eseli's Notebook*.
[27] Fernandez and Loban, *Zamiyakal*, Titui (p. 45), Southern Cross (p. 42), The Tagai (p. 51), and Comet (p. 48).

Straits). The dying person's spirit climbs to the tallest tree and is launched across the sky. If the fireball fragments, the falling 'sparks' (*uir uir*, pronounced wier-wier) tell the observer that the person died leaving a large family behind. The 'booming' sound the fireball makes when it explodes is called *dum* (doom), which is associated with a large, eight-person drum.

**Timing of Dances**

Another important astronomical element of some Islander dances is that they are held at astronomically significant times. Each year a ceremonial dance is performed on the sacred islet of Pulu off the western coast of Mabuyag. The dance is performed to remember the dead and to celebrate a time of abundance. The timing of the ceremony signifies the yams are ready to be harvested, the coming of the dry season, and the arrival of the cool southeast trade winds. It notifies the people that tubers will increase in size, the rainbow bird (*Merops ornatus*) will migrate north, and the turtles will begin mating. This is ceremony is timed by the rising of the yam-star, *Kek* (Arcturus), which rises at dusk in late April.[28]

**Concluding Remarks**

Astronomical knowledge and symbolism is prominent in Torres Strait Islander dances, songs, and musical material culture. This knowledge is encoded in the song, the dance performance, and the timing of dances, serving as a mnemonic for retaining culture and passing knowledge to successive generations. Astronomically inspired dances, headdresses, and dance machines remain a vital and important component of Islander musical traditions.

**Acknowledgements**

We acknowledge Torres Strait Islander elders and communities, for this is their knowledge and their intellectual property. I thank to Martin Nakata, Doug Passi, Leah Lui-Chivizhe, Michael Passi, Elsa Day, Aaron Bon, Jim Nai, Tommy Pau, and anonymous participants. This project was funded by the Australian Research Council (DE140101600) and the ethnographic fieldwork was approved by the Monash Human Research Ethics Committee (HC15035).

**Bibliography**

Beckett, Jeremy. *Torres Strait Islanders: custom and colonialism*. Cambridge: Cambridge University Press, 1987.
Bosun, David. *Zugubau Mabaig.* Australian Art Network (2007a).

---

[28] Eseli, "*Eseli's Notebook*. pp. 20-1.

http://australianartnetwork.com.au/shop/artwork/zugubau-mabaig/.
Bosun, David. *Merlpal Mari Pathanu.* Australian Art Network (2007b). http://australianartnetwork.com.au/shop/artwork/merlpal-mari-pathanu/.
Eseli, Peter. *Eseli's Notebook.* Edited by Anna Shnukal and Rod Mitchell. University of Queensland: Aboriginal and Torres Strait Islander Studies Unit Research Report Series, Vol. 3, 1998.
Fernandez, Robyn, and Emma Loban. *Zamiyakal: Torres Strait dance machines.* Thursday Island: Gab Titui Cultural Centre, 2009.
Haddon, Alfred C. *Cambridge anthropological expedition to the Torres Straits,* 6 vols. Cambridge: Cambridge University Press, 1901-1935.
Hamacher, Duane W. 'Frameworks for studying cultural astronomy in the Torres Strait.' *Journal of Australian Anthropology,* in review.
Hamacher, Duane W. and Ray P. Norris. 'Eclipses in Australian Aboriginal Astronomy.' *Journal of Astronomical History and Heritage* 14, no. 2 (2011): pp. 103-14.
Karttunen, Hannu, Pekka Kröger, Heikki Oja, Markku Poutanen, and Karl J. Donner. *Fundamental Astronomy.* Berlin, Heidelberg, New York: Springer, 2007.
Kepa, W., N. Pegrum, and K. Nuenfeldt. *Lagau Kompass: Music & Dance From St. Paul's Community (Mua Island) Torres Strait.* Cairns: Pegasus Studios, 2013.
Leaman, T. M., D. W. Hamacher, and M. Carter. 'Aboriginal Astronomical Traditions from Ooldea, South Australia, Part 2: Animals in the Ooldean Sky'. *Journal of Astronomical History and Heritage* 19, no. 1 (2016) ): pp. 61-78.
Nakata, Martin. 'The cultural interface of Islander and Scientific knowledge'. *Australian Journal of Indigenous Education* 39 (Supplement) (2010): pp. 53-7.
Sharp, Nonie. *Stars of Tagai.* Canberra: Aboriginal Studies Press, 1993.
Whop, John. 'Stories Under Tagai' (video) Brisbane: State Library of Queensland, 2012. https://www.youtube.com/watch?v=5kU4EvV9yI8.
Wilson, Lindsay. *Kerkar Lu: contemporary artefacts of the Torres Strait Islanders.* Brisbane: Queensland Department of Education, 1993.

**Duane Hamacher** is a Senior Research Fellow at the Monash University Indigenous Centre in Melbourne.
**Alo Tapim** is an Indigenous Torres Strait Islander and Meriam elder from Mer (Murray Island), Queensland.
**Segar Passi** is an Indigenous Torres Strait Islander and Meriam elder from Mer (Murray Island), Queensland.
**John Barsa** is an Indigenous Torres Strait Islander and Meriam elder from Mer (Murray Island), Queensland.